\documentclass[
superscriptaddress,
 amsmath,amssymb,
 aps,
 pra,
 twocolumn,
 10pt,
 longbibliography
]{revtex4-2}

\usepackage{algorithm}
\usepackage[noend]{algpseudocode}

\usepackage{orcidlink}
\usepackage[utf8]{inputenc}
\usepackage[T1]{fontenc}
\usepackage{graphicx}
\usepackage{array}

\usepackage{booktabs}
\usepackage{multirow}
\usepackage{color,soul}
\usepackage{mathtools}
\usepackage{hyperref}
\hypersetup{
	colorlinks,
	linkcolor={blue},
	citecolor={blue},
	urlcolor={blue}
}
\usepackage[capitalise]{cleveref}
\usepackage[braket]{qcircuit}

\usepackage{xcolor}

\renewcommand{\selectlanguage}[1]{}

\begin{document}

\title{Scalable decoding protocols for fast transversal logic in the surface code}

\author{Mark L. Turner}
\email{mark.turner@riverlane.com}
\affiliation{Riverlane, Cambridge, CB2 3BZ, UK}
\author{Earl T. Campbell}
\affiliation{Riverlane, Cambridge, CB2 3BZ, UK}
\affiliation{School of Mathematical and Physical Sciences, University of Sheffield, Sheffield S3 7RH, UK}
\author{Ophelia Crawford}
\affiliation{Riverlane, Cambridge, CB2 3BZ, UK}
\author{Neil I. Gillespie}
\affiliation{Riverlane, Cambridge, CB2 3BZ, UK}
\author{Joan Camps}
\affiliation{Riverlane, Cambridge, CB2 3BZ, UK}

\date{May 2025}

\begin{abstract}
Atomic, molecular and optical (AMO) approaches to quantum computing are promising due to their increased connectivity, long coherence times and apparent scalability. However, they have a significantly reduced cadence of syndrome extraction compared to superconducting devices, a potentially crippling slow-down given the substantial logical gate counts required for quantum advantage.
Transversal logic, which exploits higher connectivity, has the potential to significantly speed up the logical clock rate by reducing the number of syndrome extraction rounds required, but current decoders for fast transversal logic are not scalable. This is not just because existing decoders are too slow to handle the large decoding volumes resulting from fast logic; transversal logic breaks the key structural properties that make real-time decoding of lattice surgery efficient.
We introduce two new, windowed decoding protocols for transversal logic in the surface code that restore modularity and locality to the decoding problem.
Using our protocols, we show that, with a very small space overhead, our scalable decoders unlock an order of magnitude speed-up for transversal logic compared to lattice surgery.
Taken together, our results provide key evidence for the viability of large-scale algorithms on AMO qubits.
\end{abstract}

\maketitle

\section{Introduction}

Atomic, molecular and optical (AMO) approaches to quantum computing, notably neutral atom and trapped ion devices, have seen impressive experimental progress, culminating in demonstrations of foundational primitives of logic across numerous logical qubits~\cite{bluvstein_logical_2024, rodriguez_experimental_2024, reichardt_logical_2024, paetznick_demonstration_2024}. However, despite many favourable characteristics, slower physical gates in AMO systems lead to Quantum Error Correction (QEC) cycle times in the order of ms~\cite{poole_architecture_2025, an_high_2022}, compared to $\mu$s in their superconducting counterparts~\cite{acharya_quantum_2025, krinner_realizing_2022, besedin_realizing_2025, caune_demonstrating_2024, ali_reducing_2024}. Without significant progress in accelerating AMO systems, the tractability, and therefore viability, of large-scale fault tolerant algorithms on such hardware is at risk. While improvements are to be expected in physical gate times, they need to be complemented by QEC strategies that decisively improve the logical clock speed of the quantum computer.

This could be possible as AMO systems offer qubit shuttling and connectivity that is daunting to replicate in superconducting systems. Transversal gates, which exploit increased connectivity to operate directly on data qubits to implement a fault tolerant logical gate, are a promising way to accelerate these platforms. For example, Cain et al.~\cite{cain_correlated_2024} show that consecutive Clifford gates can be executed with $O(1)$ rounds of syndrome extraction between each gate by configuring a hypergraph decoder~\cite{delfosse_toward_2021} with visibility across all logical qubits in the system. In an analogous lattice surgery context, these same circuits would be substantially more costly, because each gate layer would require $O(d)$ rounds of syndrome extraction~\cite{horsman_surface_2012}.

\begin{figure*}[t!]
\centering
\includegraphics[clip, trim=1.5cm 18.8cm 15.2cm 0.6cm, width=500pt]{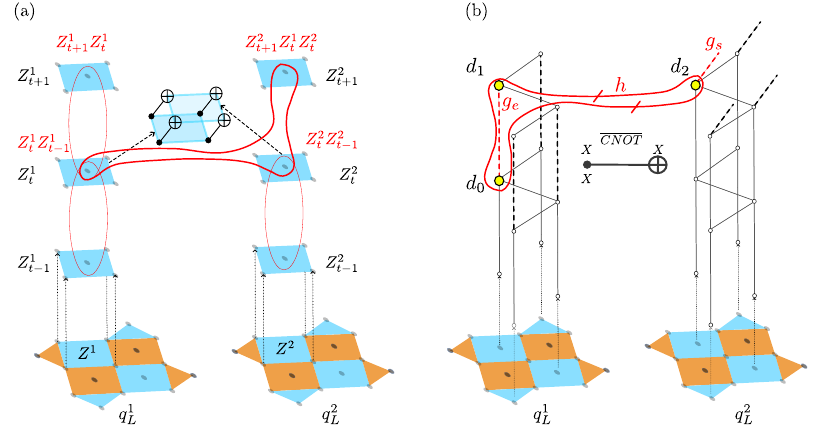}
\caption{~\textbf{Construction of detectors for transversal entangling gates.}~\textbf{(a)} 
Here we show $Z$ stabiliser plaquettes from two separate qubits being measured at three time steps with a transversal CNOT immediately after time step $t$. Combinations of measurements that form detectors are circled in red.
The transversal CNOT copies errors between logical qubits---$X$ ($Z$) errors from control (target) to target (control)---broadening the sensitivity of certain stabiliser measurements to errors from across the system.
To maintain the determinism of detector definitions and spacetime locality in error detection, the detector defined at time $t+1$ on $Z^2$ is computed to be the parity of three measurements: $Z^2_{t+1} Z^1_t Z^2_t$~\cite{gidney_stim_2021}. This accounts for the possibility of $X$ errors on $q^1_L$ propagating through the transversal CNOT and flipping $Z^2_{t+1}$.~\textbf{(b)} A measurement error affecting the stabiliser measurement at $Z_t^1$ will therefore flip three detectors, $d_0=Z^1_t Z^1_{t-1}$, $d_1=Z^1_{t+1} Z^1_t$, $d_2=Z_{t+1}^2 Z_t^1 Z^2_t$, yielding the hyperedge $h = \{d_0, d_1, d_2\}$ circled in red. Note, $h$ is not trivially decomposable into a graphlike representation~\cite{cain_correlated_2024}. Decomposing $h$ into two~\textit{ghost edges} ($g_e$ and $g_s$), edges that are fragments of higher order hyperedges and unsupported by lower order terms in the detector error model, is an effective way to break down error mechanisms of this type into a more favourable form once combined with the new correlated decoding techniques described in this paper.}
\label{fig:error_model_for_transv_entangling_gates}
\end{figure*}

However, decoding fast transversal logic is a fundamentally different problem to lattice surgery where the decoding problem is well understood~\cite{bombin_modular_2023}. The scaling of the decoding volume under fast transversal logic risks overwhelming the decoder and effectively halting logical computation~\cite{sahay_error_2025}, ruling out previously proposed global hypergraph decoding techniques. In part, this problem results from the inapplicability of standard windowing schemes to the spatial dimension of fast logic, pointing to a need for new windowing protocols tailored to the unique setting of fast transversal logic. Additionally, improving decoding throughput is not the only area where progress is needed: the modular structural properties of lattice surgery that respond more favourably to real-time decoding at scale need to be recovered with innovations at the decoding level in the very different context of fast transversal logic. Finally, every time a new decoding scheme is proposed, its impact on the time complexity of transversal logic needs to be assessed.  Where previous work has focused on Clifford logic, this analysis must extend into the non-Clifford setting, and it remains the case that sustaining and scaling error correction in this context is challenging.

In this work, we address these decoding challenges in the following ways. We introduce the \textit{ghost protocol}---a decoding protocol that guarantees small, modular decoding volumes in the context of fast transversal logic in the surface code by reformulating the spatial windowing problem as a sparse message passing algorithm. Equipped with a scalable decoder, we study its performance in the context of both deep Clifford and non-Clifford logic, focusing on the structural properties of the $T$ gate realised via teleportation. We argue that in the context of fast, windowed transversal Cliffords, the time overhead of state teleportation is fundamentally a problem of buffer size. From here we illuminate the surprising inherent resilience of state teleportation to small, constant size buffer regions when constituted via transversal entangling gates. We characterise how this resilience interacts with the necessity of spatial windowing and extend it with a further decoding protocol we call \textit{patience} to improve accuracy close to what is observed with global decoding. With general applicability to Clifford+T computation~\cite{bravyi_universal_2005}, our work provides further evidence that transversal logic can be executed in $O(1)$ time. This delivers at least an order of magnitude speed-up in the logical clock rate of Clifford+T under standard lattice surgery assumptions~\cite{chamberland_universal_2022} and can be achieved with a far smaller space overhead than would be encountered with time optimal lattice surgery~\cite{fowler_time-optimal_2013}.

\textit{Note added:} as we were finalising our paper, related work was published by the Delft~\cite{serra-peralta_decoding_2025} and Harvard~\cite{cain_fast_2025} groups.

\begin{figure}[t!]
\centering
\includegraphics[clip, trim=11.8cm 18cm 0cm 0cm, width=245pt]{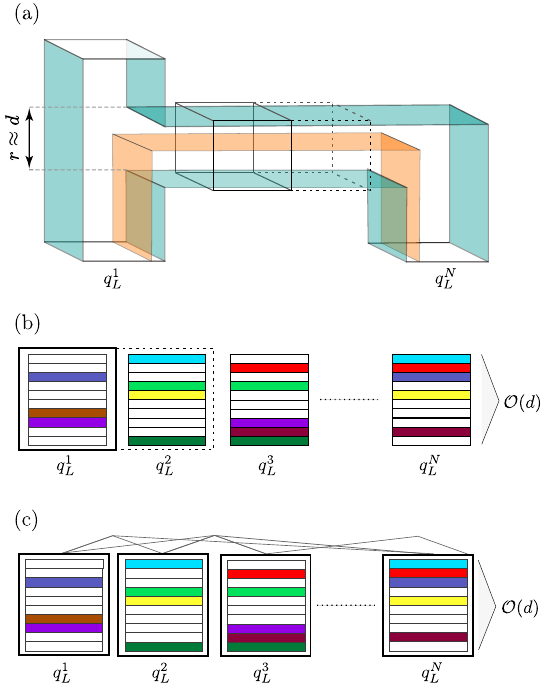}
\caption{~\textbf{Decoding volumes and the spatial windowing problem.} Windowing schemes typically work by complementing a commit region with a locally adjacent buffer region~\cite{dennis_topological_2002}. The assumption is, therefore, that there is a coherent notion of spatially local syndrome information that enables successful partial decoding. Since lattice surgery conforms to  two-dimensional nearest neighbour connectivity, this assumption holds and commit-buffer windowing techniques apply naturally~\cite{bombin_modular_2023}.~\textbf{(a)} Consequently, we can slide a decoding window (solid and dashed boxes represent the commit and buffer regions, respectively) across a merge patch of arbitrary size. Conversely, fast transversal logic denies us locality in the DEM.~\textbf{(b, c)} Here we represent $N$ logical qubits with time progressing vertically; each nested rectangle represents a round of syndrome extraction and two logical qubits sharing a rectangle of the same colour denotes that they have undergone a transversal entangling gate during this round. $O(1)$ logic is in tension with the fact decoders need $O(d)$ rounds of syndrome information to infer error patterns reliably, because it leads commit-buffer decoding to fail: as shown in~\textbf{(b)}, the solution to $q^1_L$ either directly or indirectly depends on all other qubits in the system and a buffer region around $q^2_L$ (dashed) provides insufficient additional information to reliably commit to a solution on $q^1_L$.~\textbf{(c)} We propose that logical qubits can be decoded independently if spatial windowing is rephrased as a sparse message passing problem (see Sec.~\ref{sec:ghost_protocol}).}
\label{fig:windowing_and_decoding_volumes}
\end{figure}

\section{Decoding volumes under fast logic}

Transversal gates are not universal~\cite{eastin_restrictions_2009} but nevertheless emerge as useful constructs for executing logic due to their inherent fault tolerance and low depth. A gate is \textit{transversal} if, after a partitioning of all physical qubits into disjoint parts, the physical gates used to implement the transversal gate have a tensor product structure with respect to these parts~\cite{jochym-oconnor_disjointness_2018}. The transversal logical Clifford gate set was recently completed in the rotated surface code with the observation that a fold-transversal $S$ gate could be implemented on the half-cycle state while the code's instantaneous stabilisers resemble those of the unrotated surface code~\cite{chen_transversal_2024}. The transversal $H$ can be implemented by applying physical Hadamards on all data qubits followed by a 90 degree code block rotation, escaping the need for a more involved circuit~\cite{geher_error-corrected_2024} when qubits cannot be freely shuttled. Finally, two-qubit entanglement can be achieved via physical two-qubit gates between data qubits in each respective part of two code blocks. The transversal CNOT has been demonstrated by performing a parallel move to interlace the data qubits of two logical qubits and executing physical CNOTs between each pair of data qubits that span code blocks~\cite{bluvstein_logical_2024}.

\textit{Decoders} are classical algorithms that use measurement data derived from \textit{syndrome extraction} circuitry to identify errors that occurred during a quantum computation. Transversal gates interleave with this circuitry which repeatedly measures Pauli operators called \textit{stabilisers} to enable error detection. Decoders typically consume measurement data in the form of \textit{syndromes}---sets of \textit{detector}~\cite{gidney_stim_2021} values. Detectors are parities of measurements that are deterministic under noiseless execution and get flipped from their expected values by errors.
To predict the effect of errors on logical operators, the decoder leverages syndrome data and a \textit{detector error model} (DEM): a representation of the defect patterns associated with each separate error and their effects on the logical state~\cite{gidney_stim_2021}. These representations often begin as \textit{hypergraphs} where certain single errors flip more than two detectors, though some codes and circuitry may allow for decomposition of these hypergraphs into a \textit{graphlike} representation over which the minimum weight perfect matching problem can be solved exactly in polynomial time~\cite{edmonds_paths_1965}. The Union-Find decoder~\cite{delfosse_almost-linear_2021} approximates the matching algorithm, and has seen hardware implementations with throughputs that scale sublinearly in the code distance~\cite{ziad_local_2024,liyanage_fpga-based_2024}.

The \textit{decoding volume} is defined as the number of detectors party to a given decoding problem.
When operating at scale and in real-time, the decoding problem must be divisible into smaller parts using a \textit{windowing protocol}, so it can be solved efficiently in response to a continuous stream of measurement data~\cite{dennis_topological_2002}. Crucially, the decoder reconfiguration overhead implicit in this operation must be manageable within reasonable time frames appropriate to the qubit modality.
The impact of transversal entangling gates on the decoding volume is a byproduct of measurement errors that immediately precede the round in which a transversal CNOT is performed. These errors flip three detectors and the resultant hyperedges fuse the components of the DEM associated with independent logical qubits into interdependent decoding problems (Fig.~\ref{fig:error_model_for_transv_entangling_gates}). While transversal gates preserve the locality of errors within surface code patches---and therefore fault tolerance---they also delocalise syndrome information across space, stretching the scope of the decoding problem considerably.

Strong numerical evidence has emerged that deep Clifford logic can be sustained with a very small, constant number of syndrome extraction rounds between logical gates~\cite{cain_correlated_2024}. This result has assumed correlated hypergraph decoding with full visibility across all logical qubits, an approach that neglects the necessity of windowing. The impact of fast transversal entangling gates on the decoding volume is substantial and leads Sahay et al.\ to reject the viability of $O(1)$ rounds of syndrome extraction between transversal gates~\cite{sahay_error_2025}. Since windowed decoders require $O(d)$ rounds of syndrome information to decode errors of weight $\lfloor d / 2 \rfloor$, they observe that transversal entangling gates separated by $O(1) \ll d$ rounds of syndrome extraction yields a decoding volume scaling as $2^{O(d)}$. This upper bound is derived from a worst-case ``binary-tree'' circuit that spreads an error across logical qubits exponentially in the logical circuit's depth. They argue that this exponential decoding volume could overwhelm the decoder, risking the backlog problem~\cite{terhal_quantum_2015} where the computation grinds to a halt. We note that large, albeit not exponential, spatial volumes can also be accumulated in lattice surgery via multiqubit measurements between spatially estranged patches. However, unlike in lattice surgery, fast transversal logic provides no guarantee of locality in interactions between qubits and undermines the notion of a contained buffer region, meaning that existing windowing schemes~\cite{skoric_parallel_2023, dennis_topological_2002, bombin_modular_2023, wu_fusion_nodate} used to break down decoding problems in space do not apply (Fig.~\ref{fig:windowing_and_decoding_volumes}a,b).

To solve this, Sahay et al.\ advocate enforcing $d$ rounds of syndrome extraction between transversal entangling gates so that each logical operation can be decoded in independent windows of duration $O(d)$---a time cost that slow AMO platforms are unlikely to be able to afford in the large-scale fault tolerant regime. Alternatively, one could stick to the global correlated decoding strategies of Cain et al.\ and argue that binary tree circuits do not characterise the kind of logical subroutines that might be employed in a realistic quantum algorithm. Indeed, gadgets such as Toffoli ladders, prominent in quantum arithmetic~\cite{gidney_halving_2018}, do not grow the decoding volume exponentially in the duration of the window; instead, the decoding volume would grow linearly across qubits in the temporal depth of the window given $O(1)$ transversal logic.

While such a decoding problem might be solvable with the kHz throughput necessary for AMO (from a QEC cycle duration of around 1ms), it is less obvious that this would work in a streaming context since DEMs would have limited commonality between windows, requiring the decoding problem to be reformulated and the decoder to be completely reconfigured between windows. These are additional costs independent of the decoder's throughput and could be substantial. The reason for this is that fast transversal logic forgoes the modular structure~\cite{bombin_modular_2023} that characterises decoding problems based on lattice surgery---and it is this modular structure that minimises the reconfiguration overhead that lattice surgery computation necessitates. It follows that we need to rethink how, in the context of fast transversal logic, we maintain a constant and manageable decoding volume while minimising reconfiguration between windows. This means minimising the extent to which the decoding problem changes in relation to the underlying logical circuit being decoded. With this in place, we will be better equipped to present an analysis of the time complexities of transversal gates.

\section{The ghost protocol}
\label{sec:ghost_protocol}

Our forthcoming analysis of Clifford and non-Clifford logic is founded on the observation that the decoding volume under fast transversal logic can be managed via decomposition of hyperedges that span logical qubits. Formally, these hyperedges are undecomposable since they are unsupported by lower order error mechanisms in the DEM~\cite{cain_correlated_2024}. Nevertheless, decomposition is possible: we call edges that emerge from decomposition without support through lower order terms \textit{ghost edges} (dashed edges in Fig.~\ref{fig:error_model_for_transv_entangling_gates}b) and leverage them as correlated fragments of higher order effects.

The goal of our decomposition is to exploit the concept of ghost edges to break down error mechanisms that span logical qubits so that logical qubits can be decoded independently. This means ensuring that ghost edges only flip detectors associated with one logical qubit. The hyperedges that emerge from transversal entangling gates have order three and span logical qubits so we decompose them into a time/hook-like ghost edge and a ghost singleton (an edge to the boundary that flips just one detector) in the other patch. These are represented by $g_e$ and $g_s$, respectively in  Fig.~\ref{fig:error_model_for_transv_entangling_gates}b. That is, in the DEM we artificially reinstate an edgelike time component that is otherwise subsumed into a set of hyperedges. In related work, the order-three hyperedges that emerge under the transversal $S$ gate are decomposed in a similar way~\cite{chen_transversal_2024}, though the hyperedges are intrapatch error mechanisms that span stabiliser types, not, as in our case, interpatch errors that affect the decoding volume. Our approach maintains a spatial separation between logical qubits at the level of the decomposition of the hypergraph, not through detector definitions that break the temporal locality of defect information with respect to each independent error mechanism (see~\cite{wan_iterative_2024} for a decoder that assumes detector definitions of this form).

The \textit{ghost protocol} is an iterative decoding procedure where each logical qubit is decoded to correction edges independently several times, each time exchanging messages with decoders that are responsible for other logical qubits in the system. These messages handle correlations between ghost edges and, by extension, logical qubits. Since we know ahead of time that $g_e$ is necessarily correlated with $g_s$, any decoders that decode to $g_e$ will: 1) refine their syndrome such that the defects at either end of $g_e$ are flipped; 2) message the decoder responsible for $g_s$ with the instruction to flip the single defect represented by this ghost edge. 
We track the effect of the changing syndrome on the logical operators that the decoder is trying to protect. 
Taken together, these syndrome refinement operations have the effect of committing to order-three hyperedges and eliminating them from decoding problem before subsequent iterations in the protocol are performed. Initial decoding passes therefore function like a predecoder that purges order-three hyperedges from the system and the derivation of a final logical effect is postponed until the predecoding routine has completed.
The correlations between ghost edges are highly structured, predictable and learnable ahead of time. 

The inner decoder used in our protocol is two pass correlated matching~\cite{fowler_optimal_2013}, an implementation based on~\cite{higgott_sparse_2025}. That is, we employ two levels of message passing, each of which addresses thematically distinct contributions to the error hypergraph---two pass for $Y$ errors that correlate $X$ and $Z$ errors, ghost decoding for interpatch hyperedges that correlate instances of $g_e$ and $g_s$. These two forms of correlated decoding are closely related: in both cases correlations between correction edges are used to emulate hypergraph decoding but use simpler graph-based primitives. However, correlated decoding for $Y$ errors does not revolve around the concept of ghost edges since $Y$ errors in the surface code are trivially decomposable into lower order terms.
Further, $Y$ errors are handled with two decoding passes, whereas the ghost protocol often needs to iterate more often (App.~\ref{app:parameters}): refining the syndrome can unlock the discovery of new ghost edges indicative of further order-three hyperedges that were undiscoverable through graph-based decoding of the initial syndrome.

\begin{figure}[t!]
\centering
\includegraphics[clip, trim=5.1cm 1.6cm 33.6cm 4cm, width=\columnwidth]{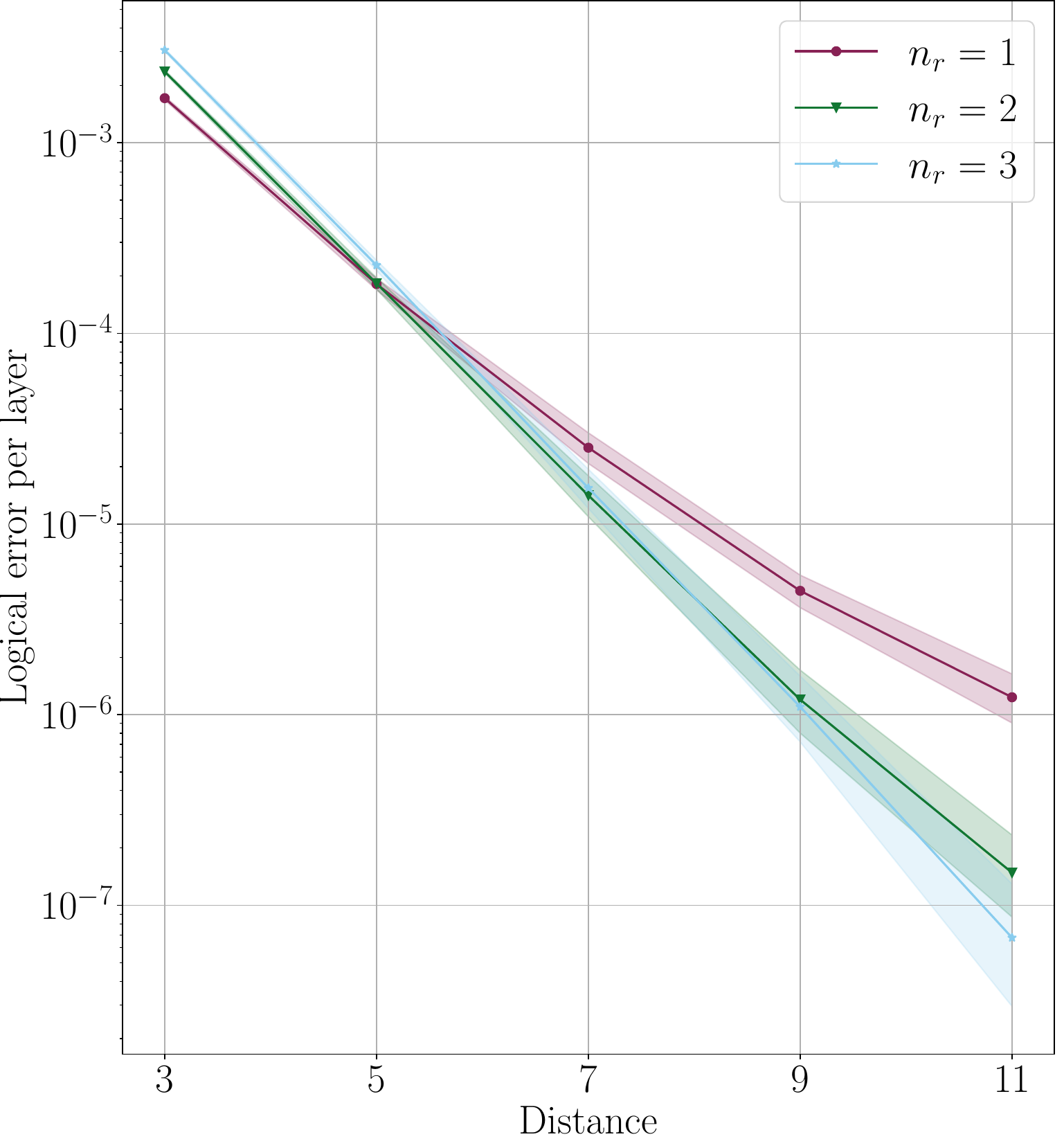}
\caption{~\textbf{Deep logical Clifford circuits.} We show the subthreshold scaling of ghost decoding of deep Clifford logic between four rotated surface codes under circuit level noise ($p=0.1\%$). Following~\cite{cain_correlated_2024}, each logical qubit is prepared in $\ket{+}$ and subject to 32 layers of transversal gates, followed by perfect measurements of the logical stabilisers of the state. A gate layer is composed of $n_r$ rounds of syndrome extraction, preceded by a randomly chosen single qubit transversal gate $\{\mathrm{H}, X, Y, Z\}$ on each qubit then two transversal CNOTs that entangle random pairs of qubits. Increasing $n_r$ can steepen our protocol's gradient of error suppression in the distance, but the spacetime volume computed as $(n_r + 1)d^2$ remains optimal at $n_r=1$ for per layer error rates of $10^{-6}$. See App.~\ref{app:parameters} for the parameters that determine the decoding passes used in these numerics and App.~\ref{app:noise-model} for details of the noise model.}
\label{fig:deep_logical}
\end{figure}

In order for the ghost protocol to be successful, we must handle $g_s$ with care. These singleton edges are inevitable byproducts of our decomposition but, if exposed to the decoder continuously, make decoding impossible. This is because they materialise as edges to the boundary across the spatial horizon of the DEM at each time step where a transversal entangling gate has taken place. Used naively, these destroy the code distance as they introduce logically non-trivial loops of low, constant weight in the DEM. Therefore, one might be tempted to make the decoder blind to these error mechanisms all together. However, during initial decoding passes these singletons can have an important role to play: consecutive transversal entangling gates in opposite directions separated by $O(1)$ rounds of syndrome extraction yield a temporally proximate $g_e$ and $g_s$ that can conflict with one another, making the discovery of $g_e$, crucial for successful ghost decoding, difficult. By retaining edges to the boundary from instances of $g_s$ during a carefully selected subset of decoding passes (see App.~\ref{app:parameters}) we can break these conflicts and reliably correlate errors across logical qubits. This works because the passes in which we expose $g_s$ are used to clean up the syndrome, not derive an overall logical effect where the shortest undetectable logical operator in the DEM is crucial. This addresses a weakness of existing graph-based decoders where an inability to decode even modestly deep circuits composed of fast transversal Clifford logic has been reported~\cite{guernut_fault-tolerant_2024}.

Fig.~\ref{fig:deep_logical} shows the subthreshold scaling of ghost decoding of deep Clifford logic between four rotated surface codes under circuit level noise ($p=0.1\%$)---a benchmark first proposed in~\cite{cain_correlated_2024}. In our numerics, the best accuracy is seen with $n_r=3$ but the spacetime volume of the computation is minimised at $n_r=1$ in the regimes we study, results that are consistent with what was previously observed with global correlated decoding methods with belief-HUF~\cite{cain_correlated_2024}. Our results are the first to show deep Clifford logic where the spatial component of the decoding volume is managed.

\begin{figure*}[t!]
\centering
\includegraphics[clip, trim=0.0cm 24.4cm 10cm 0cm, width=410pt]{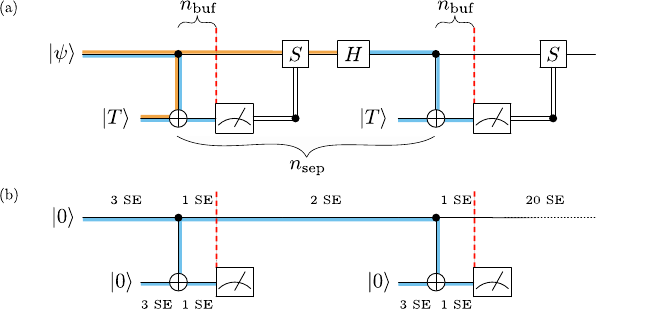}
\caption{~\textbf{$T$ gates and measures of time overhead}.~\textbf{(a)} Two consecutive $T$ gates, implemented via teleportation circuits, separated by a Hadamard. The $X$ ($Z$) logical sheet is shown in orange (blue), capturing where a logical observable is sensitive to physical $Z$ ($X$) errors. Logical sheets that do not affect a measurement are ignored for clarity. Logical measurements are made in the $Z$ basis, and each informs a classically controlled $S$ gate. During each $T$ gate, a logical $Z$ basis measurement is performed that only terminates one of the qubits---the qubit carrying the state we wish to teleport. Consequently, the act of measuring a subset of qubits in the system imposes an artificial~\textit{severance} (dashed red line) across the decoding problem associated with the top logical qubit. A severance manifests as an open temporal boundary across the error model prior and can disrupt decoding by introducing new locations for error strings to terminate without guarantees of topological protection. With $n_{\textrm{buf}}$ we quantify how many rounds of syndrome extraction separate this severance from the logical sheet (shown in blue) that needs to be decoded to close the $T$ gate. Separately, the quantity $n_{\textrm{sep}}$ captures how many rounds must separate transversal CNOTs in the context of contiguous $T$ gates while preserving robust and scalable decoding.~\textbf{(b)} A proxy of the $T$ gate circuit that is simulable in the Clifford simulator, Stim~\cite{gidney_stim_2021}. All qubits are prepared in $\ket{0}$ with $n$ SE describing how many rounds of syndrome extraction occur on each qubit at each point in the circuit. This circuit gives us an experiment to quantify $n_{\textrm{buf}}$ and $n_{\textrm{sep}}$---the quantities that determine the time complexity of $T$ gates---in the context of severances and complete windowing of fast transversal logic.}
\label{fig:T_gate}
\end{figure*}

By decomposing order-three hyperedges such that they do not span logical qubits, spatial windowing emerges naturally from the act of sparse global message passing (Fig.~\ref{fig:windowing_and_decoding_volumes}c) and the decoding volume---both in size and content---reflects a windowed single-qubit memory experiment on each decoder core regardless of the number of rounds of syndrome extraction between transversal gates. That is, the need to manage the decoding volume motivates making the problem graphlike. Further, data locality, crucial in distributed computing, is ensured despite high qubit connectivity, in that each decoder takes ownership of decoding a single logical qubit. Above all, in a streaming setting our architecture minimises the amount of reconfiguration work necessary on each decoder core between windows: from the decoder's perspective the decoding problem is simply memory-like and logic is handled via syndrome refinement gadgets that can be abstracted into a representation that is removed from the exact context in which a gadget is deployed. Consequently, the window configuration is divorced from the specific details of the logical circuitry. There is a structural predictability inherent in windows of this form that resembles the favourable modularity of real-time error correction in a lattice surgery context~\cite{bombin_modular_2023}.

It remains to study the relationship between the ghost protocol and the time complexities of \textit{non}-Clifford logic.

\begin{figure*}[t!]
\centering
\includegraphics[clip, trim=5.1cm 1.5cm 6.4cm 2.5cm, width=500pt]{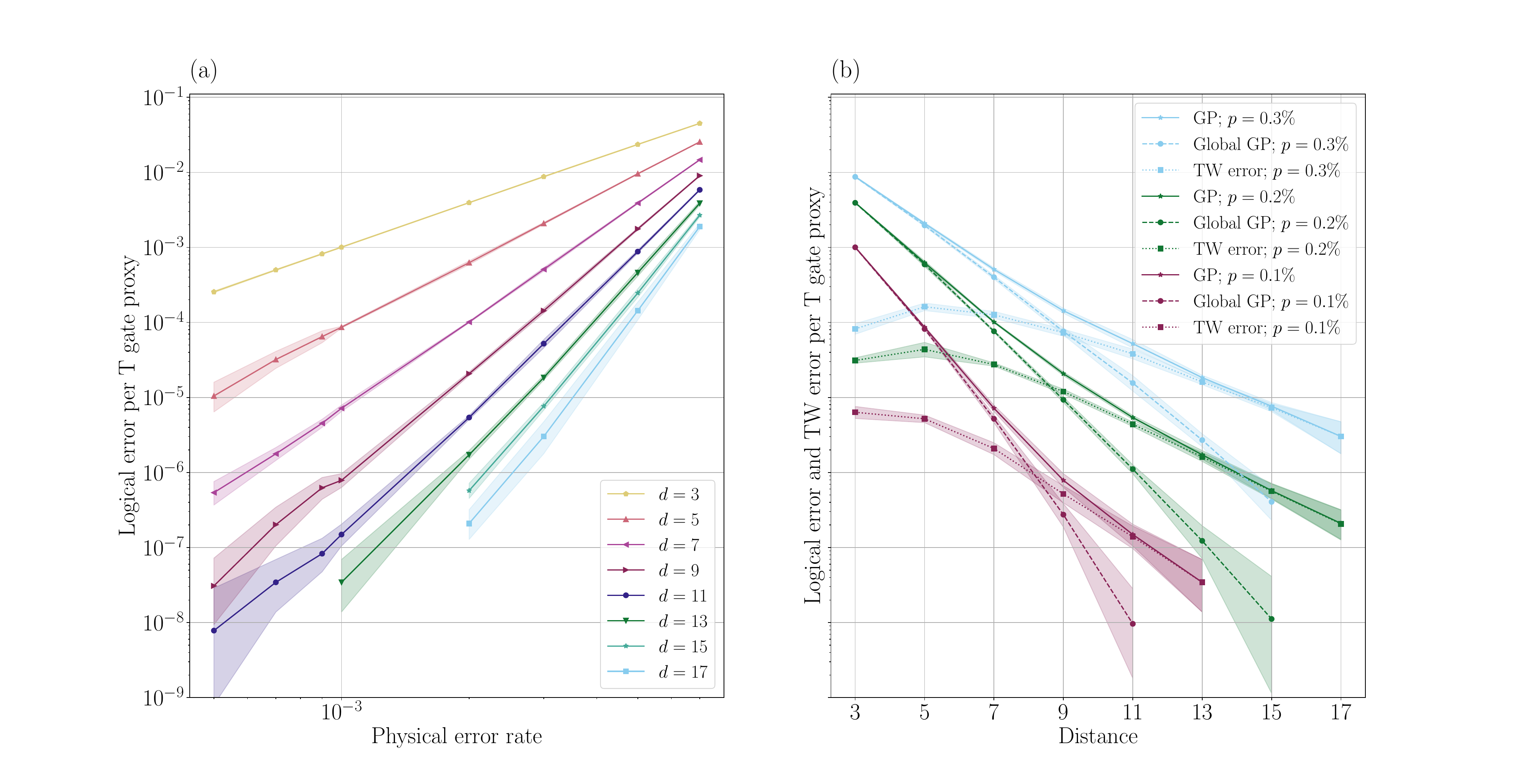}
\caption{~\textbf{Ghost decoding of $T$ gate proxies ($n_{\textrm{buf}} = 1$ and $n_{\textrm{sep}} = 3$)}.~\textbf{(a)} Logical error per $T$ gate proxy of fully windowed (i.e., both spatially and temporally) ghost decoding as a function of the physical error rate $p$.  Straight lines indicate power laws between the logical error and $p$, with the power increasing with the distance, although not quite linearly for our range of distances---note, the error mechanisms responsible for this are described in Sec.~\ref{sec:resilience} and further decoding protocols that overcome them are proposed in Sec.\ref{sec:patience}.~\textbf{(b)} Gradients of error suppression in the distance for fully windowed ghost decoding (solid lines labelled ``GP''), ghost decoding without temporal windowing (dashed lines labelled ``Global GP'', though note spatial windowing is still operational) and the temporal windowing error (dotted lines labelled ``TW error''). TW error describes the probability that Global GP disagrees with the decoding result of GP, capturing the contribution of errors originating from our aggressive temporal windowing. We note a $p$--dependent transition point (at roughly $d=11$ for our range of $p$) where the gradient of the fully windowed decoder converges on the TW error rate with a gradient of error suppression that is roughly half that of the global decoder, an asymptotic regime where we are dominated by the error mechanisms that emerge from the relationship between $n_{\textrm{buf}}$ and our spatial windowing protocol (again, this halving effect is reversed in Sec.~\ref{sec:patience} with additional decoding).
}
\label{fig:data}
\end{figure*}

\section{Fast teleportation is a buffering problem}
\label{sec:teleportation_is_buffering}

Transversal gates are not universal~\cite{eastin_restrictions_2009}, centralising the importance of gate teleportation for universal computation in the surface code. Here we focus on the $T$ gate which is performed by consuming $\ket{T} = \frac{1}{\sqrt{2}}(\ket{0} + e^{i\pi/4}\ket{1})$ states by teleportation. With reference to Fig.~\ref{fig:T_gate}, two quantities allow us to quantify the time overhead of a $T$ gate realised with transversal gates:

\begin{enumerate}
\item $n_{\textrm{buf}}$ captures the number of rounds of syndrome extraction between a transversal CNOT that creates the entanglement necessary for teleportation and the logical measurement that informs a classically controlled $S$ gate.
\item $n_{\textrm{sep}}$ defines the number of rounds that separate consecutive $T$ gates.
\end{enumerate}

Notably, $n_{\textrm{buf}}$ functions as a buffer region with respect to the decoding problem centred on the logical measurements in the circuit, presenting new structural challenges for decoding. This is because the logical measurement can be flipped by errors before the transversal CNOT on both qubits, but only on the measured qubit after the transversal CNOT. The result is a decoding problem where an open temporal boundary (``severance'') has been imposed across the surviving logical qubit, a disruptive severance that is only $n_{\textrm{buf}}$ rounds from the logical sheet we are trying to correct.

Buffer regions are typically required to be of size $O(d)$ rounds of syndrome extraction~\cite{skoric_parallel_2023, bombin_modular_2023}, suggesting $n_{\textrm{buf}}$ should scale with the code distance irrespective of the time complexity of the transversal CNOT when considered in isolation. Consequently, it would seem that in breaking free of lattice surgery, a paradigm that necessitates measuring stabilisers for $O(d)$ rounds, we are quickly forced back into such a regime by the buffering conditions of windowed error correction itself. To emphasise the point, there is a striking tension here between the practicalities of error correction and the theoretical speed at which logic can flow through the system. Therefore, the key question is: how small can we make $n_{\textrm{buf}}$ while still decoding the $Z$ logical sheet (blue in Fig.~\ref{fig:T_gate}) effectively?

To study the behaviour of the decoding protocol as a function of $n_{\textrm{buf}}$ and $n_{\textrm{sep}}$, we simulate a proxy (Fig.~\ref{fig:T_gate}b) of the repeated teleportation circuit shown in Fig.~\ref{fig:T_gate}a. Our goal here is to use a logical circuit that captures the key structural properties of teleportation circuitry while simultaneously being efficiently simulable within the Clifford simulator, Stim~\cite{gidney_stim_2021}. To achieve this we strip out classically controlled logical gates since they are non-Clifford and prepare all states in $\ket{\overline{0}}$ to avoid non-stabiliser states. Further, we modify the circuit such that our observables are deterministic. As such, the $H$ gate between the $T$ gates in Fig.~\ref{fig:T_gate}a is omitted in our proxy. In the ideal circuit $H$ enables consecutive $T$ gates to evolve the state towards an arbitrary state, however, in our context it would make the second logical $Z$ measurement nondeterministic and resistant to simulation in Stim. Instead, we leave two rounds of syndrome extraction between the
measurement of a logical qubit and 
the subsequent transversal CNOT in the expectation that a classically controlled $S$ gate would be applied midcircuit during the first round~\cite{chen_transversal_2024} and the transversal $H$ would be applied at the end of that same round of syndrome extraction. This then leaves a further empty round of syndrome extraction before the next CNOT which begins another $T$ gate transformation. Ultimately, this gives us a fast $T$ gate proxy where $n_{\textrm{sep}} = 3$ and a benchmark that focusses on the open temporal boundaries we call ``severances'' (dashed red lines in Fig.~\ref{fig:T_gate}) that challenge the decoder when logical gates are executed in $O(1)$ time.

Our proxy does not address the time complexities of state preparation around which there is an extensive literature~\cite{dennis_topological_2002, bravyi_lieb-robinson_2006, hastings_topological_2011, gidney_inplace_2024, wan_constant-time_2024, bravyi_universal_2005, litinski_magic_2019, gidney_efficient_2019, gidney_magic_2024}. Instead, our focus is the time overhead of teleportation in the bulk of the computation. In Fig.~\ref{fig:data} we show the surprising result that $n_{\textrm{buf}} = 1$ and $n_{\textrm{sep}}=3$ yields error suppression even with full temporal and spatial windowing.  This suppression scales with $\lceil d / 2 \rceil$ at small distances, and $\lceil d / 4 \rceil$ at higher distances. This is also the point where errors stemming from temporal windowing dominate. Temporal windowing (``TW'') errors are shown as the dotted line in~\ref{fig:data}b, capturing the probability of the temporally windowed ghost protocol disagreeing with the global ghost protocol (note, this metric is equivalent to first step heralded errors in~\cite{zhou_algorithmic_2024}). In the next section, we illuminate why temporal windowing errors come to dominate and characterise transition points in the observed gradients of error suppression. We relate this to the unique structural properties of the post-transversal-entangling-gate setting. This analysis culminates in a new decoding strategy---\textit{patient decoding}---that recovers the accuracy lost at higher distances under aggressive temporal windowing with negligible time overhead.

\section{Inherent resilience with limits}\label{sec:resilience}

In this section we will elucidate the error patterns that determine the characteristics of the error suppression observed in Fig.~\ref{fig:data}b and explain why it persists despite our aggressive temporal windowing of the decoding problem.

\begin{figure}[t!]
\centering
\includegraphics[clip, trim=0.1cm 21.3cm 10.5cm 0cm, width=\columnwidth]{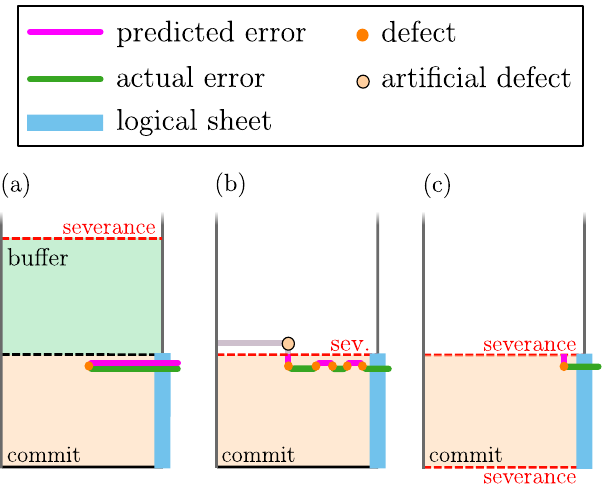}
\caption{~\textbf{Basics of buffer size for quantum memories}.~\textbf{(a)} The maximum weight ($w = \lfloor d/2 \rfloor$) adversarial error string that is correctable in a sliding window setting~\cite{dennis_topological_2002}, given a buffer region of size $\approx d$.~\textbf{(b)} Given a buffer region of size 0, the maximum weight correctable adversarial error shrinks to $w = \lfloor d/4 \rfloor$. Here we show an error pattern described in~\cite{bombin_modular_2023} that can occur when this is exceeded. Note, error suppression is still possible in absence of a buffer region, because the effect of a buffer region is imitated by the act of pushing artificial defects into the future.~\textbf{(c)} Conversely, if forgoing the use of artificial defects but retaining a buffer region of size 0, the decoder would be susceptible to small, constant weight logical errors. That is, increased independence between decoding windows is incompatible with minimal buffer region sizing.
}
\label{fig:windowed-memory}
\end{figure}

First, however, we reflect on the behaviour of error suppression in a quantum memory context under various windowing protocols---a frame of reference for the problem to hand. A typical temporal windowing for a decoding problem contains ``commit'' and ``buffer'' regions.  The decoder addresses the whole window, but it only decides firmly on outcomes inside the commit regions. The buffer region expands the view of the decoder beyond the commit region and generally includes around $d$ rounds of syndrome information. This enables reliable correction of all errors up to weight $\lfloor d / 2 \rfloor$ (Fig. \ref{fig:windowed-memory}a) and the expected logical error suppression behaviour, $P_L \propto \Lambda^{- \lceil d / 2 \rceil}$. Importantly, if we utilise the sliding window protocol described in~\cite{dennis_topological_2002}, error suppression in the distance persists even if no buffer region is used. In this context, the minimum weight of uncorrectable errors is diminished from~$\approx d/2$ to~$\approx d/4$ (Fig.~\ref{fig:windowed-memory}b) halving the gradient of error suppression in the distance. Error suppression is nevertheless possible here, because the sliding window protocol leverages artificial defects to postpone decisions around the logical effect of ``ambiguous'' error patterns into the future---an effect that functions to imitate something of a buffer region even where, formally, no such buffer region exists.

\begin{figure}[t!]
\centering
\includegraphics[clip, trim=0cm 0cm 10.5cm 9cm, width=250pt]{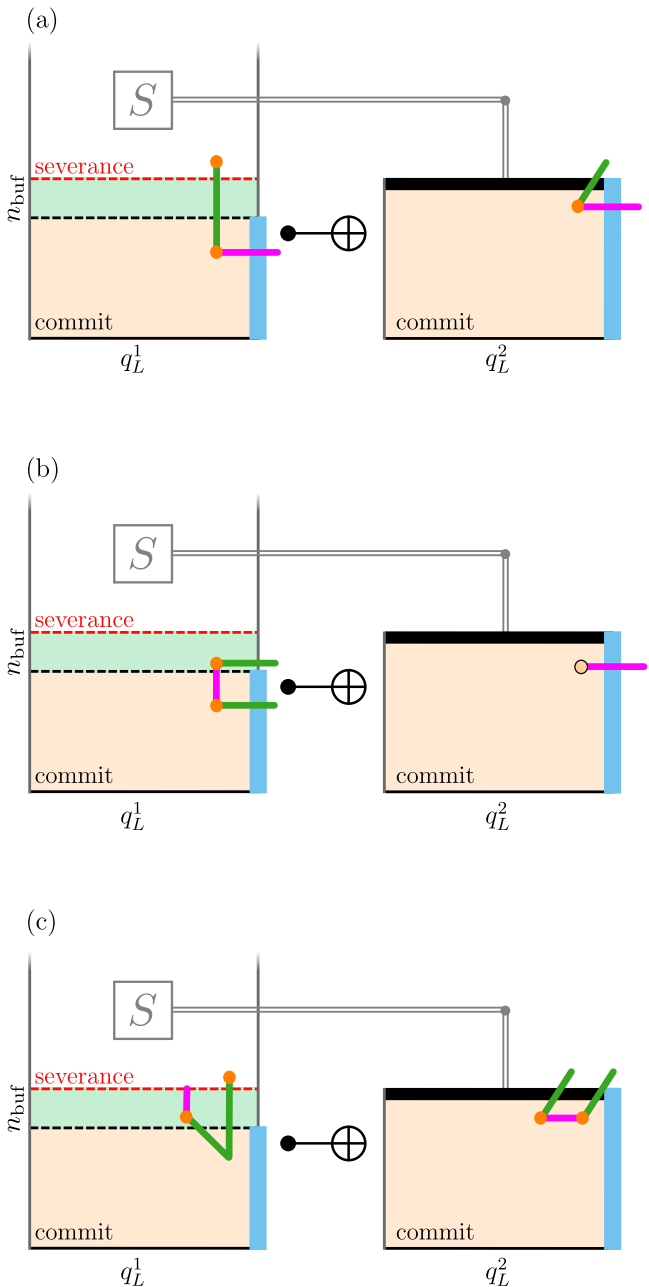}
\caption{~\textbf{Resilience of the ghost protocol to $T$ gates with $n_{\textrm{buf}} = 1$}.~\textbf{(a)}
We observe that error strings that pierce through entangling gates and terminate after a severance activate logically trivial loops in the error hypergraph. Here the decoder never infers that an order-three hyperedge occurs, but nevertheless arrives at a logically equivalent solution by crossing the logical sheet separately on both $q_L^1$ and $q_L^2$.~\textbf{(b)} Similarly, the effect of spatial errors that lead the decoder to erroneously derive that an order-three hyperedge \textit{did} occur is neutralised by a further correction string introduced on $q_L^2$ in response to a syndrome refinement event (note the new artificial defect on $q_L^2$).~\textbf{(c)} More than one order-three hyperedge can occur and be undiscovered without jeopardising the logical information.
}
\label{fig:resilient-windowed-t-gates}
\end{figure}

Conversely, a sliding window protocol that does not operate artificial defects will, in the absence of a buffer region, fail to suppress errors in the distance (Fig.~\ref{fig:windowed-memory}c). Structurally, this is more akin to the temporal windowing problem that emerges in the context of fast $T$ gates: we cannot allow ourselves to benefit from the use of artificial defects that bind a firm decoding decision to a future decoding problem since a final logical effect is needed as soon as possible to apply a classically controlled gate. To do otherwise would return the time complexity of the $T$ gate to $O(d)$. Therefore, against the backdrop of windowed quantum memory, a surprising feature of our data in Fig.~\ref{fig:data}b is that strong error suppression nevertheless persists despite $n_{\textrm{buf}} = 1$. So what is different here and how is this possible given the $O(1)$ time complexity of our $T$ gate proxy?

Here we have to return to what is special about the DEMs that underpin transversal entangling gates (recall Fig.~\ref{fig:error_model_for_transv_entangling_gates}b). The core difference in comparison to the memory setting, is that transversal entangling gates create \textit{more} defect information than standard errors in the surface code. In particular, error strings that begin before a transversal entangling gate and terminate after it deterministically spread defect information onto other qubits due to the subsumption of timelike errors into order-three hyperedges that link the otherwise detached error models of two logical qubits. That is, error strings fork in space due to the copying of errors through transversal gates.

Fig.~\ref{fig:resilient-windowed-t-gates} gives a topological perspective on why this increased syndrome information sustains error correction even when this same syndrome information is incomplete in proximity to a commit region. We show how the copying of errors through transversal gates opens up a wealth of logically trivial loops in the hypergraph that the ghost protocol is able to exploit when decoding to logical effects. Notably, failure to discover a ghost edge and trigger syndrome refinement (Fig.~\ref{fig:resilient-windowed-t-gates}a,c)---or conversely, erroneous discovery of a ghost edge (Fig.~\ref{fig:resilient-windowed-t-gates}b)---is \textit{not} inevitably a failure mode. This underlies the viability of fast $T$ gates with aggressive windowing.

\begin{figure}[t!]
\centering
\includegraphics[clip, trim=11.2cm 0cm 0cm 9.6cm, width=245pt]{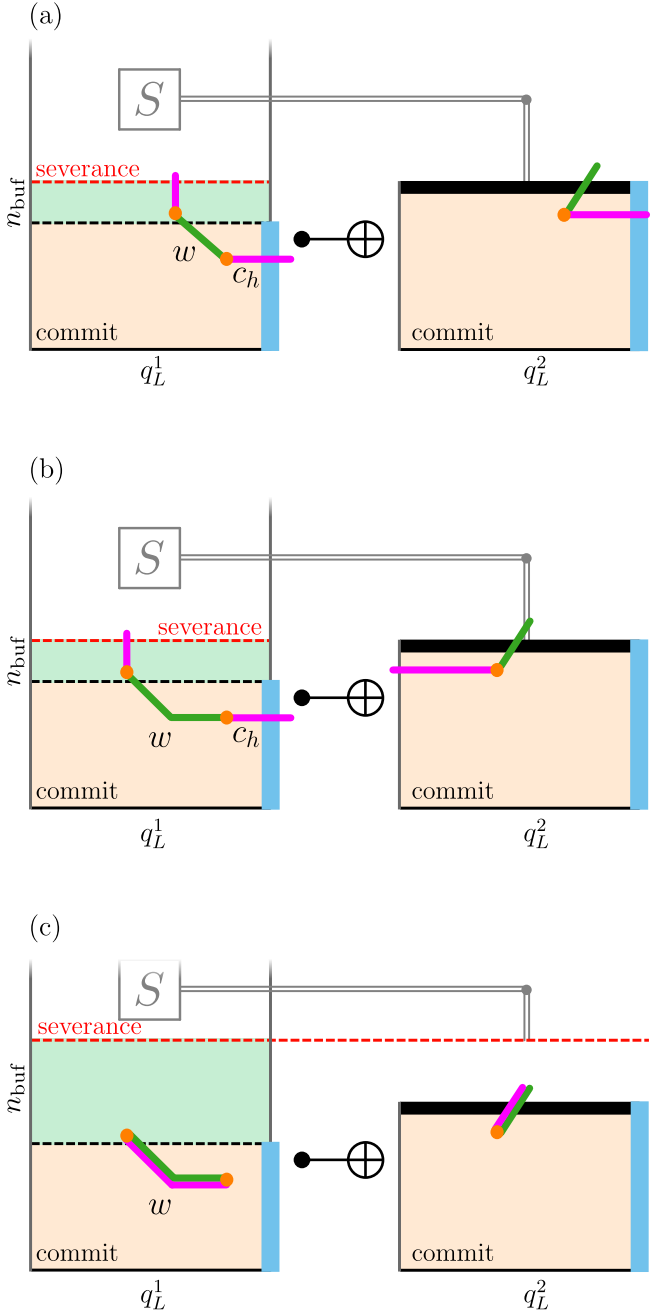}
\caption{~\textbf{Limits of the ghost protocol's resilience given $T$ gates with $n_{\textrm{buf}} = 1$}.~\textbf{(a)} Here we show an error of weight $w$ and a windowed correction of weight $c=c_h+n_{\textrm{buf}}+1$ in qubit $q_L^1$. A logical error is avoided but only because $w$ is small.~\textbf{(b)} Error strings that are offset from a boundary by $c_h$ and travel across and up through an entangling gate by $w$ are uncorrectable if $w + c_h \geq \lceil d/2 \rceil$, transforming (a) into a failure mode. Here a logical error is introduced because the actual error does not flip the logical, but unlike before, 
the effect of $c_h$ crossing the logical in $q_L^1$ is no longer neutralised in $q_L^2$ due to the increase in $w$.~\textbf{(c)} The naive fix to the aforementioned failure mode is to enforce $n_{\textrm{buf}} \approx d/2$. Sec.~\ref{sec:patience} improves upon this.
}
\label{fig:limits-of-resilient-windowed-t-gates}
\end{figure}

These unusual topological structures create redundancy and enable the error suppression in the distance shown numerically in Fig~\ref{fig:data}. However, there are limits to this resilience that fall short of the ordinary level of protection we expect from our code distance. We find that the ghost protocol is subject to logical errors of weight (see Fig.~\ref{fig:limits-of-resilient-windowed-t-gates}):
\begin{equation}\label{eq:min-failure-weight}
    w\geq\frac{n_{\textrm{buf}}+1+\lceil d/2\rceil}{2}\,.
\end{equation}
We have verified by search that the smallest error patterns leading to failure in our decoder saturate this inequality.

By comparing Fig.~\ref{fig:limits-of-resilient-windowed-t-gates}a and~\ref{fig:limits-of-resilient-windowed-t-gates}b we show how we can transition into one of these failure modes. Logical errors are introduced when the concatenation of the error and correction strings extends past the middle of $q_L^1$, $w+c_h\geq\lceil d/2\rceil$, which we can rewrite as $w+c\geq n_{\textrm{buf}}+1+\lceil d/2\rceil$ using that the total correction weight is $c=c_h+n_{\textrm{buf}}+1$ ($+1$ for the edge to the open temporal boundary). The decoder will pick the purple correction in $q_L^1$ when $w\geq c$. Adding the two inequalities, $c$ cancels and we deduce Eq.~\eqref{eq:min-failure-weight}. Crucially, logical errors are suppressed in the distance despite $n_{\textrm{buf}} = 1$, a significant difference to the aforementioned windowing of memory without artificial defects. However, there is a halving of the effective distance that is implied by this inequality. This halving is slow to affect the gradients of error suppression we showed numerically in Fig.~\ref{fig:data}b, because $n_{\textrm{buf}} + 1$ acts as a constant in Eq.~\eqref{eq:min-failure-weight} that inflates $w$, creating the illusion of error suppression scaling with $\lceil d / 2 \rceil$ at low $d$. At intermediate $d$ the impact of this constant is diminished and error suppression scales with $\lceil d / 4 \rceil$, evidencing the halving of the effective distance anticipated in Eq.~\eqref{eq:min-failure-weight}.

Notably, this failure mode is a feature of our windowing protocol where we consider logical qubits separately. Decoders that forgo spatial windowing~\cite{cain_correlated_2024, sahay_error_2025} achieve the expected code distance as they do not suffer from this failure mode: given the error in Fig.~\ref{fig:limits-of-resilient-windowed-t-gates}b, decoders with visibility over both $q_L^1$ and $q_L^2$ can use the defect information on $q_L^2$ to infer the actual error string on $q_L^1$. Consequently, in a regime where decoders are not spatially windowed, the resilience property we outline here survives without limitations even with $n_{\textrm{buf}} = O(1)$---a key difference to what is possible in normal temporal windowing settings.

Instead, to guarantee ordinary protection against all errors up to weight $\lfloor d/2 \rfloor$ it follows from Eq.~\eqref{eq:min-failure-weight} that ghost decoding needs a buffer size of $n_{\textrm{buf}} \approx d/2$ (Fig.~\ref{fig:limits-of-resilient-windowed-t-gates}c)---a big cost for transforming the problem into the windowed form advocated for in Sec.~\ref{sec:ghost_protocol}. Without increasing $n_{\textrm{buf}}$ in this way, ghost decoding would still enable $O(1)$ logic, but with a fourfold space overhead in the number of qubits. In the following section we overcome this with a new decoding strategy that allows $O(1)$ transversal logic and windowed decoding to coexist with a significantly reduced accuracy penalty.

\section{All you need is patience}\label{sec:patience}

Here we describe a new decoding technique---\emph{patience}---that builds on the ghost protocol and seeks to overcome the aforementioned distance reducing failure modes with minimal impact on $n_{\textrm{buf}}$. We begin by highlighting two important properties fundamental to patience:

\begin{enumerate}
    \item The freedom to postpone the application of conditional logic during teleportation. That is, we are free to accumulate more syndrome information than originally planned for via further syndrome extraction rounds before applying a conditional gate.
    \item The inherent rarity of the distance reducing failure modes described in Sec.~\ref{sec:resilience}. This rarity is evident in the low TW error probability observed in Fig.~\ref{fig:data}b.
\end{enumerate}

Property (1) means $n_{\textrm{buf}}$ can be a runtime parameter without changing the underlying logical operation (in the absence of error, a $T$ gate remains a $T$ gate regardless of how many rounds separate a CNOT from the classically controlled $S$ gate) and property (2) is suggestive of the fact that increases to $n_{\textrm{buf}}$ are only necessary in a heavily suppressed number of cases.

\begin{figure}[t!]
\centering
\includegraphics[clip, trim=5cm 1.7cm 33.6cm 4cm, width=\columnwidth]{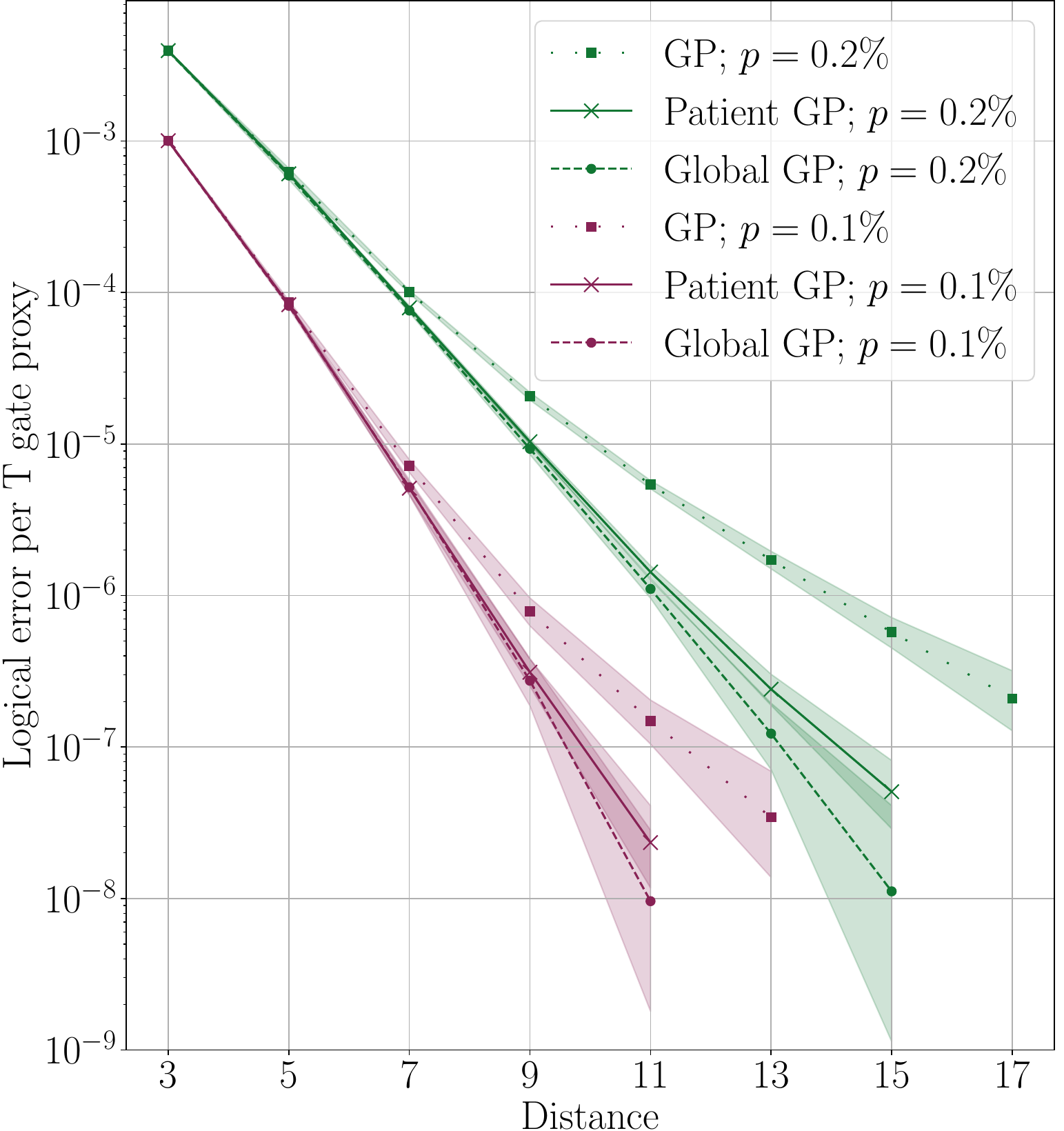}
\caption{~\textbf{Patient temporal windowing decoder}. Here we overlay the accuracy data presented in Fig.~\ref{fig:data} with further data derived from when the ghost protocol is extended with patience (``Patient GP''). When the decoder heralds the need to be patient, we collect further rounds of syndrome information before re-decoding the logical measurement. In a fully windowed setting, patience overcomes the failure modes of weight $\approx d/4$ described in Sec.~\ref{sec:resilience}. This is achieved without incurring a costly time penalty from enforcing $n_{\textrm{buf}} \approx d/2$ across all $T$ gates. See Table~\ref{tab:patience} for data on the average time penalty of patience.
}
\label{fig:patience}
\end{figure}

Patience is a protocol that aims to exploit these two properties in order to escape the failure modes that afflict the ghost protocol without necessitating a significant increase in \textit{average} $n_{\textrm{buf}}$. Specifically, patience extends the ghost protocol with the ability to herald distance reducing failure mechanisms resulting from aggressive temporal windowing. In the minority of cases where the decoder raises a herald, the protocol postpones the application of a conditional gate until $n_{\textrm{buf}}$ has increased through continued syndrome extraction rounds. This means increasing $n_{\textrm{buf}}$ such that $w=\lceil d / 2 \rceil$ where $w$ is defined in Eq.~\ref{eq:min-failure-weight}. Once further syndrome data has been generated and passed to the decoder, the decoder has a final attempt at determining a correction with more syndrome information.

Operationally, we trigger patience through two inference steps either of which is sufficient to immediately trigger patience if satisfied:

\begin{enumerate}
    \item  We run the ghost protocol as usual, but track whether the weight of the correction string increases between the first and final decoding iterations performed by the ghost protocol. This functions to detect most cases where aggressive temporal windowing spuriously triggers a syndrome refinement operation (of the form described in Sec.~\ref{sec:ghost_protocol}) despite the fact no such order-three hyperedge occurred.  In the language of Fig.~\ref{fig:error_model_for_transv_entangling_gates}b, these cases are often detectable because they mistakenly involve a flip of $g_s$, triggering a defect that, in future decoding passes, incurs a long set of correction edges due to the string-like nature of errors in the surface code. The change in syndrome weight is assessed in a local region so that the analysis is not distracted by distant parts of the decoding problem. In the particular case of our $T$ gate proxy, this means we localise this analysis on the target qubit in temporal proximity to the severance (dashed red line in~\ref{fig:T_gate}b)---the place where increases to the weight of the correction string is most directly indicative of a likely temporal windowing failure.
    \item Inspired by soft-output decoding constructions~\cite{gidney_yoked_2023, meister_efficient_2024}, we perform a further complementary matching on a modified DEM, where the open temporal boundary is turned into a closed boundary, meaning error strings can no longer terminate at a temporal boundary. With this complementary matching, we can herald cases that yield a different correction bit to the previous decoding with the open temporal boundary. This functions to detect most cases where the ghost protocol fails to discover order-three hyperedges that occurred (for example, the case shown in Fig.~\ref{fig:limits-of-resilient-windowed-t-gates}b).
\end{enumerate}

\begin{table}[]
\setlength{\tabcolsep}{4pt}
\renewcommand{\arraystretch}{1.5}
\begin{tabular}{@{}llll}
\toprule
\multirow{2}{*}{\textbf{$d$}} & \multirow{2}{*}{\textbf{delay}} & \multicolumn{2}{c}{\textbf{Average patience delay (num rounds)}} \\
& &  $p=0.001$ & $p=0.002$ \\
\cmidrule{1-4}
3 & 0 & 0 & 0 \\
5 & 1 & $(5.227 \pm 0.051) \times 10^{-4}$ & $(2.284 \pm 0.014) \times 10^{-3} $ \\
7 & 2 & $(1.125 \pm 0.007) \times 10^{-3} $ & $(5.031 \pm 0.020) \times 10^{-3}$ \\
9 & 3 & $(2.324 \pm 0.005) \times 10^{-3}$ & $(9.764 \pm 0.015) \times 10^{-3}$ \\
11 & 4 & $(4.029 \pm 0.006) \times 10^{-3}$ & $(1.545 \pm 0.001) \times 10^{-2}$ \\
13 & 5 & -  & $(2.077 \pm 0.001) \times 10^{-2}$ \\
15 & 6 & -  & $(2.492 \pm 0.002) \times 10^{-2}$ \\
\cmidrule{1-4}
\end{tabular}
\caption{
\label{tab:patience}
\textbf{Average patience delay}. 
For each distance and noise model simulated in Fig.~\ref{fig:patience}, we report the number of rounds of syndrome extraction performed in $n_{\textrm{buf}}$ when patience is applied (see ``delay''). With these patience delays, $n_{\textrm{buf}}$ is increased such that $w=\lceil d / 2 \rceil$ where $w$ is defined in Eq.~\eqref{eq:min-failure-weight}. Without a patience delay $n_{\textrm{buf}}=1$ is fixed. Since patience is applied rarely 
in the subthreshold regime, its impact on average $n_{\textrm{buf}}$ per $T$ gate proxy is negligible. Recall that without patience, ghost decoding of state teleportation in the large $d$ regime would require an overhead of $n_{\textrm{buf}} \approx d/2$ per $T$ gate to reliably match the error suppression behaviour of the global decoder. We note that reductions in the physical error rate $p$ minimise the overhead of patience.
}
\end{table}

Together, these techniques give us a mechanism for efficiently reducing the gap between the error suppression seen under global decoding of $T$ gates and the temporally windowed decoder (Fig.~\ref{fig:patience}). Patience is effective because TW errors are rare (see Fig.~\ref{fig:data}b) and therefore to herald them (even with a relatively high probability of false positives) and delay the computation when these events are encountered has a minimal impact on average $n_{\textrm{buf}}$ and, by extension, the logical clock rate of the quantum computer. False positive heralds are benign in terms of accuracy, though the goal is to suppress them such that $n_{\textrm{buf}}$ remains small on average without increasing false negatives, which inevitably lead to logical errors. Table~\ref{tab:patience} confirms that the average cost of patience is negligible in the subthreshold regime. Together, this confirms the viability of fast $T$ gates in a fully windowed context.

\section{Discussion and outlook}

In this work we proposed two new decoding protocols---the ghost protocol and patience---that combine to enable scalable decoding of fast transversal logic. We analysed the performance of these protocols in the context of both Clifford and non-Clifford logic with a focus on the structural properties of the $T$ gate realised via teleportation. Further, we characterised the topological structures that give rise to a resilience property that facilitates fast state teleportation despite aggressive temporal windowing.

By lattice surgery, a multiqubit $\pi/8$ rotation requires $2d$ rounds of syndrome extraction on a planar lattice using the Litinski architecture~\cite{litinski_game_2019} due to the complexity of measuring arbitrary Pauli products~\cite{chamberland_universal_2022}. Achieving upwards of $10^9$ T/Toffoli gates by lattice surgery with kHz syndrome extraction circuitry has daunting implications for the time overhead of large-scale fault tolerant algorithms. By enabling scalable, fast transversal logic our protocols deliver in excess of an order of magnitude speed-up for qubits with high connectivity. This is achieved without the significant increases in the space overhead of the computation incurred in~\cite{fowler_time-optimal_2013, litinski_active_2022}. The topic of fast transversal logic is subtle and expert opinion is divided, so future non-Clifford simulations that study the time complexity of logic in relation to specific decoding schemes may be valuable.

Compressing the spatial dimension of a quantum computation is a developed and active area of research~\cite{tillich_quantum_2009, breuckmann_quantum_2021, panteleev_degenerate_2021, bravyi_high-threshold_2024, shaw_lowering_2025, berthusen_toward_2025, malcolm_computing_2025, xu_constant-overhead_2024, poole_architecture_2025}. While our study concerns the surface code, which has a lower encoding rate and therefore greater space overhead than non-surface qLDPC codes, it is important to consider the comparative value of compressing different axes of the spacetime volume in a way that accounts for the specificities of the underlying hardware. In an AMO context, the speed of syndrome extraction means we anticipate that codes with favourable characteristics in time may be preferable even over codes that reduce the spacetime volume more substantially through acting in space. Further study will be necessary to quantify this trade-off, but we suggest that the possibility of fast transversal logic in the surface code preserves its competitiveness even against the backdrop of further innovation in the qLDPC space. Regarding the applicability of our decoding protocols to the broader qLDPC context, we note that our protocols are not dependent on graph-based decoders and could be coupled with fast hypergraph decoders~\cite{wolanski_ambiguity_2025, hillmann_localized_2024, yin_symbreak_2024} that will be necessary in this area.

Fast transversal logic is difficult to decode not just from a decoder throughput perspective, but also because it challenges key structural properties that make real-time decoding of lattice surgery efficient. In lattice surgery we have a modular, block-like structure which gives predictability and repetitiveness to the decoding problem as well as a local memory model that benefits highly distributed computation. Under fast transversal logic it would seem like we have to completely reformulate the decoding problem from scratch in response to each logical measurement. With patient ghost decoding, fast transversal logic is windowed with minimal reconfiguration overhead. Our protocols enable fast transversal logic at scale, securing a crucial speed-up for qubit types with high connectivity but slow physical gate times.

\vspace{1cm}

\section*{Acknowledgements}

We thank Ben Barber, Dan Browne and Joonas Majaniemi for feedback on a draft of this paper. We thank Luka Skoric for valuable discussions about windowing. We thank Steve Brierley and our colleagues at Riverlane for creating a stimulating environment for research.

\section*{Author contributions}

M.L.T led the research, developing, implementing and evaluating the decoding protocols. M.L.T and J.C. wrote the paper with input from all authors.  All authors contributed to discussions on research direction and problem solving approaches.  O.C. and M.L.T. generated the circuits simulated in this work.

\bibliography{references_final}

\appendix

\section{Noise model} \label{app:noise-model}

We adopt the circuit level noise model proposed in~\cite{barber_real-time_2025} across all our numerical simulations. The model is parameterised by a single probability $p$ and introduces independent channels of the following type and strength throughout our Stim circuits:
\begin{itemize}
    \item Depolarisation of both qubits after each two-qubit Clifford gate with probability $p$.
    \item Depolarisation of each idle qubit and after each single qubit Clifford gate, including measurement and reset operations, with probability $p/10$.
    \item Randomly change the result of a measurement with probability $p$.
\end{itemize}
The only exception to this is that logical stabilisers at the end of our deep logical circuits (Fig.~\ref{fig:deep_logical}) are measured noiselessly with multi-qubit Pauli product measurements. All error bars in our data represent likelihoods within a factor of 1000 of the max likelihood hypothesis, given the sampled data. All Stim circuits simulated in this paper will be available on Zenodo (DOI: 10.5281/zenodo.15544576).

\section{Parameters and decoding passes} \label{app:parameters}

By default the ghost protocol performs 4 decoding passes. That is, all logical qubits get decoded independently four times, each time exchanging messages between one another and refining the syndrome (see Sec.~\ref{sec:ghost_protocol} for a full description of the message passing). In this default configuration, the first decoding pass involves turning on ghost singletons ($g_s$ in Fig.~\ref{fig:error_model_for_transv_entangling_gates}b) in the first pass, a fast reweighting operation that is linear in the number of ancilla qubits. This default configuration is used in the numerical simulations of the $T$ gate proxy---i.e., the data shown in Fig.~\ref{fig:data} and Fig.~\ref{fig:patience}. We believe the default configuration to be sufficient for most logical circuits.

The number of decoding passes and the content of these passes (whether or not ghost singletons, $g_s$, are exposed to the decoder) is however a parameter that needs to be evaluated in relation to the structure of the logical circuit, the number of rounds of syndrome extraction between logical gates and the code distance. For the numerical simulations of deep logical circuits (Fig.~\ref{fig:deep_logical}) the default configuration underlies the data points at all values of $n_r$ for $d=3,5,7,9$. However, at $d=11$ and $n_r=2,3$ we have to increase the number of decoding passes to six, displaying ghost singletons during the first and fourth decoding pass, to maintain exponential error suppression in the distance. At $d=11$ and $n_r=1$ the protocol has to work harder still to purge the syndrome of interpatch, order-three hyperedges, performing 8 passes, showing boundary edges on the first, fourth and sixth decoding pass. We currently determine these parameters empirically and further study is needed to work out how to set these parameters automatically in regimes that are hard to simulate. A fundamental principle that has worked well is to include ghost singletons in the decoding problem during the initial pass and then occasionally in intermediate decoding iterations; for the reasons discussed in Sec.~\ref{sec:ghost_protocol} ghost singletons should never be shown during the final pass.

As the number of rounds between transversal CNOTs shrinks, order-three hyperedges offset in time and with opposite orientations conflict with one another and it takes more time to discover them. Each time we discover a ghost edge there is a possibility that a subsequent matching opens up the possibility of further ghost edges appearing in a subsequent correction string. Under ghost decoding this can only be solved iteratively. We have not seen the number of passes required to decode effectively exceed the distance of the code giving us confidence in the tractability of our scheme at high distances.

\end{document}